\documentclass[aps,preprint,floatfix,showpacs,superscriptaddress]{revtex4}

\usepackage{graphicx}
\usepackage{amssymb}
\usepackage{bm}

\newcommand{\be}{\begin{equation}}
\newcommand{\ee}{\end{equation}}
\newcommand{\bel}[1]{\be\label{#1}}
\newcommand{\re}[1]{Eq.~(\ref{#1})}

\newcommand{\ds}{\displaystyle}

\newcommand{\hsp}{\hspace*{1pt}}

\begin{document}
\title{
Mach shocks induced by partonic jets\\
in expanding quark--gluon plasma}

\author{L.M. Satarov}

\affiliation{Frankfurt Institute for Advanced Studies,
J.W.~Goethe Universit\"{a}t, Max-von-Laue-Str.~1, D--60438 Frankfurt am Main, Germany}

\affiliation{Institute f\"{u}r Theoretische Physik,
J.W.~Goethe Universit\"{a}t, Max-von-Laue-Str.~1, D--60438 Frankfurt am Main, Germany}

\affiliation{The Kurchatov Institute, Russian Research Center,
123182 Moscow, Russia}

\author{H.~St\"ocker}

\affiliation{Frankfurt Institute for Advanced Studies,
J.W.~Goethe Universit\"{a}t, Max-von-Laue-Str.~1, D--60438 Frankfurt am Main, Germany}

\affiliation{Institute f\"{u}r Theoretische Physik,
J.W.~Goethe Universit\"{a}t, Max-von-Laue-Str.~1, D--60438 Frankfurt am Main, Germany}

\author{I.N. Mishustin}

\affiliation{Frankfurt Institute for Advanced Studies,
J.W.~Goethe Universit\"{a}t, Max-von-Laue-Str.~1, D--60438 Frankfurt am Main, Germany}

\affiliation{The Kurchatov Institute, Russian Research Center,
123182 Moscow, Russia}


\begin{abstract}

We study Mach shocks generated by fast partonic jets propagating
through a deconfined strongly--interacting matter. Our main goal is to
take into account different types of collective motion during the
formation and evolution of this matter.  We predict a significant deformation of Mach
shocks in central Au+Au collisions at RHIC and LHC energies as compared
to the case of jet propagation in a static medium. The observed
broadening of the near--side two--particle correlations in
pseudorapidity space is explained by the Bjorken--like longitudinal
expansion. Three--particle correlation measurements are proposed for a more
detailed study of the Mach shock waves.

\end{abstract}


\maketitle

\section{Introduction}

Sideward peaks have been recently
observed~\cite{Adl03a,Adl03b,Wan04,Jac05} in azimuthal distributions of
secondaries associated with the high~$p_T$ hadrons in central
Au+Au collisions at \mbox{$\sqrt{s_{NN}}=200$\,GeV}. In Ref.~\cite{Sto04} such
peaks were predicted as a signature of Mach shocks created by partonic
jets propagating through a quark--gluon plasma (QGP) formed in a
heavy--ion collision. Analogous Mach shock waves were studied
previously in a cold hadronic
matter~\mbox{\cite{Hof74,Sto80,Sto86,Cha86,Ris90}} as well as in
nuclear Fermi liquids~\cite{Gla59,Kho80}. The Mach--shock induced
electron emission from metal surfaces have been predicted~\cite{Sch78}
and then observed experimentally in Ref.~\cite{Fri80}.  Recently, Mach
shocks from jets in the QGP have been studied in Ref.~\cite{Cas04} by
using a linearized fluid--dynamical approach. It has been argued in
Refs.~\cite{Sto04,Rup05} that Mach--like motions of quark--gluon matter
can appear via the excitation of collective plasmon waves by the moving
color charge associated with the leading jet.

Most of these studies were dealing with the idealized case of
homogeneous, static matter. On the other hand, collective expansion
effects are known to be strong in relativistic collisions of
heavy nuclei~\cite{Nux04}. For example, thermal fits of RHIC data give
for most central events the radial flow velocities
$v_f\sim 0.6\hsp c$\,. The flow effects accompanying jet
propagation through the expanding QGP have been considered
within different approaches in Refs.~\cite{Arm04,Gal04,Gal05}.
Numerical solutions of fluid--dynamical equations with an additional
external source have been studied in Ref.~\cite{Cha05}.

It is well known~\cite{Lan59} that a point--like perturbation (a small
body, a hadron or parton etc.) moving with a supersonic speed in the
spatially homogeneous ideal fluid produces the so--called Mach region
of the perturbed matter. In the fluid rest frame (FRF) the Mach region
has a conical shape with an opening angle with respect to the direction
of particle propagation given by the expression\hsp\footnote{
Here and below the quantities in the FRF are marked by tilde.
}
\bel{mac1}
\widetilde{\theta}_M=\sin^{-1}\left(\frac{c_s}{\widetilde{v}}\right)\,,
\ee
where $c_s$ denotes the sound velocity of the unperturbed (upstream)
fluid and $\widetilde{\bm{v}}$ is the particle velocity with respect to the
fluid. In the FRF, trajectories of fluid elements (perpendicular to the
surface of the Mach cone) are inclined at the angle
$\Delta\theta=\pi/2-\widetilde{\theta}_M$  with respect to
$\widetilde{\bm{v}}$\,.
Strictly speaking, formula (\ref{mac1}) is applicable only for weak,
sound--like perturbations. It is certainly not valid for space--time
regions close to a leading particle.  Nevertheless, we shall
use this simple expression for a qualitative analysis of flow
effects. Following Refs.~\cite{Sto04,Cas04} one can estimate the angle
of preferential emission of secondaries associated with
a fast jet in the QGP. Substituting~\footnote{
Units with $c=1$ are chosen throughout the paper.
}
$\widetilde{v}=1, c_s=1/\sqrt{3}$ into~\re{mac1}
gives the value
$\Delta\theta\simeq 0.96$\,.
This agrees well with positions of maxima
of the away--side two--particle distributions observed
in central Au+Au collisions at RHIC energies.

In this paper we study influence of flow effects
on the properties of Mach shocks created by
a high--energy parton in expanding QGP. In Sect.~II we apply
simple kinematic considerations to study
weak Mach shocks propagating in transverse and collinear
direction with respect to the flow velocity.
In Sect.~III we use the method of characteristics
to calculate the width of the perturbed region
in the rapidity space. Finally, in Sect.~IV
we summarize our results and give outlook for future studies.

\section{Deformation of Mach shocks due to radial flow}

\begin{figure*}[htb!]
\vspace*{-7cm}
\centerline{\includegraphics[width=0.7\textwidth]{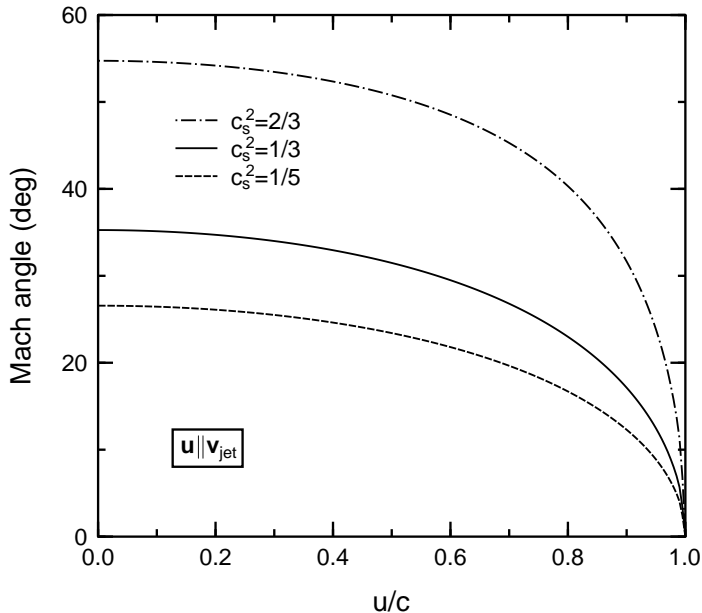}}
\caption{Mach cone angles for jet propagating collinearly to the
matter flow as a function of fluid velocity $u$\,. Different curves
correspond to different values of sound velocity $c_s$\,.}
\label{fig1}
\end{figure*}
Let us consider first the case when the away--side jet propagates with
velocity $\bm{v}$ parallel to the matter flow velocity $\bm{u}$\,.
Assuming that $\bm{u}$ does not change with space and time, and
performing the Lorentz boost to the FRF, one can see that a weak Mach
shock has a conical shape with the axis along $\bm{v}$\,. In this
reference frame, the shock front angle $\widetilde{\theta}_M$ is given
by~\re{mac1}.  Transformation from the FRF to the c.m. frame (CMF)
shows that the Mach region remains conical, but the Mach angle becomes
smaller in the CMF:
\bel{macp1}
\tan{\theta_M}=\frac{\ds 1}{\ds\gamma_u}\tan{\widetilde{\theta}_M}\,,
\ee
where $\gamma_u\equiv (1-u^2)^{-1/2}$ is the Lorentz factor
corresponding to the flow velocity $\textbf{u}$\,.
Eqs.~(\ref{mac1})--(\ref{macp1}) give the resulting expression
for the Mach angle in the CMF
\bel{macp2}
\theta_M=\tan^{-1}
\left(c_s\sqrt{\frac{1-u^2}{\widetilde{v}^{\hsp 2}-c_s^2}}\right)\,,
\ee
where
\bel{vrel}
\widetilde{v}=\frac{v\mp u}{1\mp v\hsp u}\,,
\ee
and upper (lower) sign corresponds to the jet's motion in (or opposite to)
the direction of collective flow. For ultrarelativistic jets ($v\to 1$) one
can take $\widetilde{v}\simeq 1$ which leads to a simpler expression
\bel{macp3}
\theta_M\simeq\tan^{-1}\left(\frac{\ds c_s\gamma_s}{\ds \gamma_u}\right)=
\sin^{-1}\left(c_s\sqrt{\frac{1-u^2}{1-u^2\hsp c_s^2}}\right)\,,
\ee
where $\gamma_s=(1-c_s^2)^{-1/2}$\,. According to \re{macp3},
in the ultrarelativistic limit $\theta_M$ does not depend on the 
direction of flow with respect to the jet. The Mach cone 
becomes more narrow as compared to jet propagation in static matter. This
narrowing effect has a purely relativistic origin. Indeed, the
difference between $\theta_M$ from \re{macp3} and the Mach angle in
absence of flow ($\lim\limits_{u\to 0}{\theta_M}=\sin^{-1}{c_s}$) is of
the second order in the collective velocity $u$\,. The Mach angle
calculated from~\re{macp3} is shown in Fig.~\ref{fig1} as a function of
$u$ for different sound velocities $c_s$\,. Following
Ref.~\cite{Cas04}, the value $c^2_s=1/5$ is identified
with the hadronic matter and $c^2_s=1/3$ with ideal QGP
composed of massless quarks and gluons. The value
$c^2_s=2/3$ is chosen to represent a strongly coupled QGP~\cite{Shu04}.

\begin{figure*}[htb!]
\vspace*{-7cm}
\centerline{\includegraphics[width=0.7\textwidth]{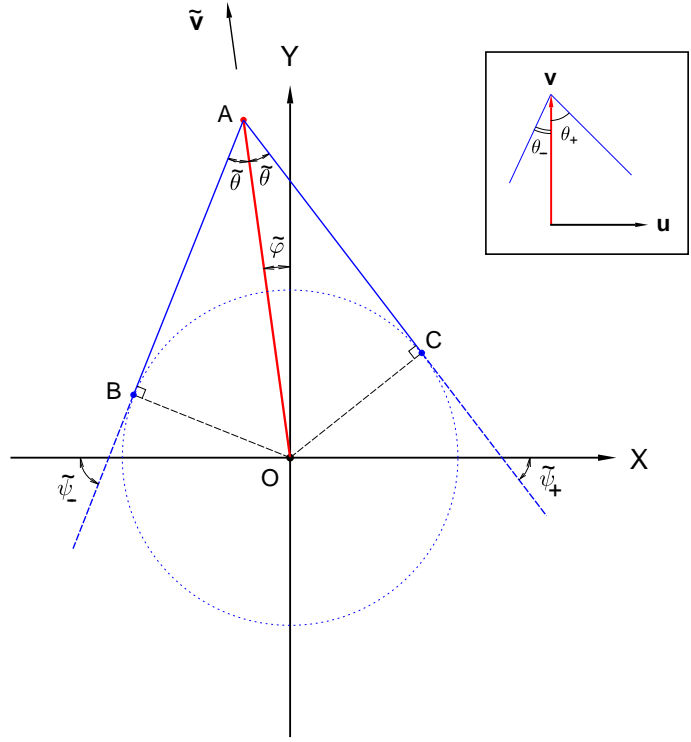}}
\caption{Mach region created by jet
moving with velocity $\bm{v}$ orthogonal to the fluid velocity~$\bm{u}$\,.
Main plot and insert correspond to FRF and CMF, respectively.
It is assumed that jet moves from $O$ to $A$ in FRF. Dotted
circle represents the front of sound wave generated at point $O$.}
\label{fig2}
\end{figure*}

The case of a jet propagating at nonzero angle with respect to the flow
velocity is more complicated. As will be shown later, Mach shocks
become nonconical for non--collinear flows. For simplicity, below we
study only the case when the jet and flow velocities are
orthogonal to each other, $\bm{v}\perp\bm{u}$\,. Let axes $OX$ and $OY$
be directed along $\bm{u}$ and $\bm{v}$\,, respectively. As before, we
first make transition to the FRF by performing the Lorentz boost along
the $OX$ axis. The components of the jet velocity $\widetilde{\bm{v}}$
in the new reference frame are equal to \bel{vtco}
\widetilde{v}_x=-u\,,\hspace*{1cm}\widetilde{v}_y=v\sqrt{1-u^2}\,.
\ee
The angle $\widetilde{\varphi}$ between vectors $\widetilde{\bm{v}}$
and $\bm{v}$ (see Fig.~\ref{fig2}) can be found from the relations
\bel{phi}
\tan{\widetilde{\varphi}}=-\frac{\ds \widetilde{v}_x}{\ds \widetilde{v}_y}=
\frac{\gamma_u u}{v}\,\raisebox{-1.2ex}{$\stackrel{\longrightarrow}
{\mbox{$\scriptstyle v\to 1$}}$}\,\raisebox{.2ex}{$\gamma_u u$}\,.
\ee

Let us now consider the situation when the jet propagates along the
path $OA=\widetilde{v}\hsp\widetilde{t}$ during the time interval
$\widetilde{t}$ in the FRF, as illustrated in Fig.~\ref{fig2}. At
the same time the wave front from a point--like perturbation (created
at point $O$) reaches a spherical surface with radius
$OB=OC=c_s\widetilde{t}$\,. Two tangent lines $AB$ and $AC$ show
boundaries of the Mach region\hsp\footnote{
Such region exists only if $\widetilde{v}>c_s$\,. The latter condition
is fulfilled if $v>c_s$ or $u>c_s$ hold.
}
with the symmetry axis $OA$\,. This region has a conical shape with
opening angles $\widetilde{\theta}$ determined by the expressions
(cf.~\re{mac1})
\bel{mcan}
\sin{\widetilde{\theta}}=\frac{OC}{OA}=\frac{\ds c_s}{\ds\widetilde{v}}
\simeq{c_s}\,.
\ee

Performing inverse transformation from FRF to CMF, it is easy to show
that the Mach region is modified in two ways. First, it is no longer
symmetrical with respect to the jet trajectory in the CMF. The insert
in Fig.~\ref{fig2} shows that the boundaries of Mach wave have
different angles, $\theta_+\neq\theta_-$, with respect to $\bm{v}$ in
this reference frame. One can interpret this effect as a consequence of
transverse flow which acts like a ''wind'' deforming the Mach cone
along the direction $OX$. On the other hand, the angles of the Mach
front with respect to the beam ($OZ$) axis are not changed under the
transformation to CMF. We conclude, that due to effects of
transverse flow, the Mach region in the CMF should have a shape of
a deformed cone with an elliptic base.
\begin{figure*}[htb!]
\vspace*{-7cm}
\centerline{\includegraphics[width=0.7\textwidth]{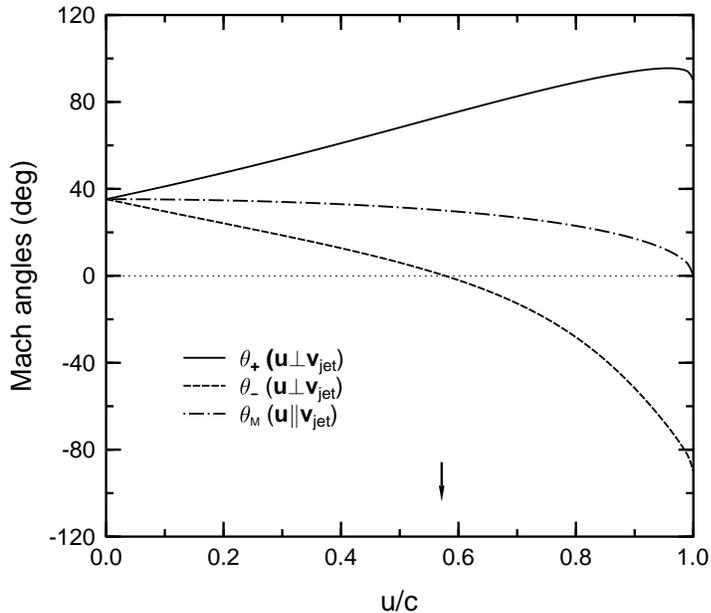}}
\caption{Angles of Mach region created by a jet
moving transversely (solid and dashed curves) and
collinearly (dashed--dotted line) to the fluid
velocity~$\bm{u}$\, in the CMF. All curves correspond to the
case~\mbox{$c_s^2=1/3$}\,. Arrow marks the value $u=c_s$\,.
}
\label{fig3}
\end{figure*}

To find the Mach angles $\theta_\pm$ it is useful
to introduce the angles $\widetilde{\psi}_\pm$  of the boundary lines
$AB$ and $AC$ with respect to the $OX$ axis in the FRF (see Fig.~\ref{fig2}).
Under the Lorentz boost to CMF, $\widetilde{\psi}_\pm$ are transformed
to angles $\psi_\pm$ which can be found by using the relations
\bel{psia}
\cot{\psi_\pm}=\gamma_u\cot{\widetilde{\psi}_\pm}=\gamma_u
\tan{(\widetilde{\theta}\pm\widetilde{\varphi})}\,.
\ee
The final expressions for Mach angles $\theta_\pm=\pi/2-\psi_\pm$
take the form
\bel{thpm}
\tan{\theta_\pm}=\cot{\psi_\pm}\simeq\gamma_u
\frac{\ds \gamma_s\hsp c_s\pm\gamma_u\hsp u}
{\ds 1\mp\gamma_s\hsp c_s\gamma_u\hsp u}\,.
\ee
The last equality gives the approximate formula
in the ultrarelativistic case $v\simeq 1$\hsp\footnote{
This formula is not accurate if either $u$ or $c_s$ are close to
unity.
}.
According to~\re{thpm}, $\theta_-$ becomes negative for supersonic
flows, i.e. at $u>c_s$\,. For such strong flows, the jet trajectory
lies outside the Mach region. One can see that the difference of the
Mach angles $\theta_\pm$ in moving and static matter is
approximately linear in $u$\,. Figure~\ref{fig3} shows the numerical
values of the Mach angles for an ultrarelativistic jet moving through
the QGP transversely or collinearly to its flow velocity. The solid and
dashed curves are calculated by using~\re{thpm} with
$c_s=1/\sqrt{3}$\,. We point out a much stronger sensitivity of the
Mach angles $\theta_\pm$ to the transverse flow velocity
as compared with the collinear flow.

\begin{figure*}[htb!]
\vspace*{-8cm}
\hspace*{3cm}\includegraphics[width=0.8\textwidth]{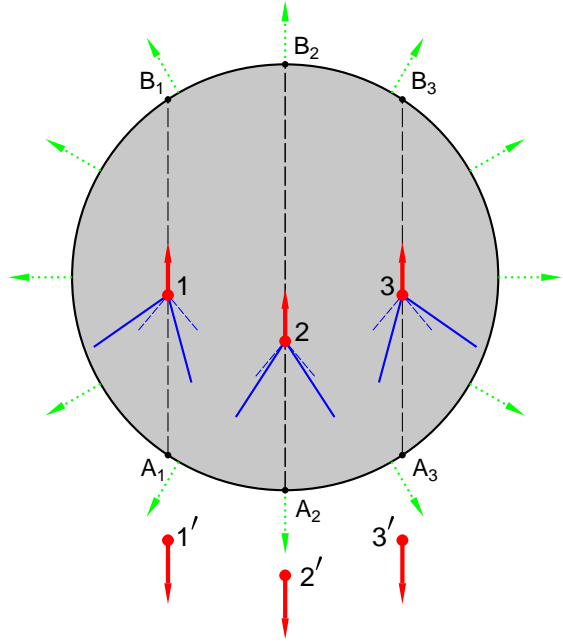}
\vspace*{-1cm}
\caption{Schematic picture of Mach shocks from jets
$1,2,3$ propagating through the fireball matter (shaded circle)
created in a central heavy--ion collision.
Dotted arrows represent local velocities of fireball expansion.
Thick downward arrows show associated trigger jets. The Mach shock boundaries
are shown by solid lines. Short--dashed lines give positions of shock
fronts in the case of static fireball.}
\label{fig4}
\end{figure*}
To discuss possible observable effects, in Fig.~\ref{fig4}
we schematically show events with different di--jet
axes $A_iB_i\,(i=1,2,3)$\, with respect to the center of a fireball
\hsp\footnote{
For simplicity we consider the case when both trigger ($i^\prime$)
and away--side ($i$) jets have zero pseudorapidities in CMF.
}.
In the $2-2^\prime$ event, the away--side jet '2' propagates along the
diameter $A_2B_2$\,, i.e. collinearly with respect to the collective
flow. In the two other cases, the di--jet axes are oriented along the chords,
$A_1B_1$ and $A_3B_3$\,, respectively. In such events, the fluid
velocity has both transverse and collinear components with respect to
the jet axis. In Fig.~\ref{fig4} we also show how the
Mach fronts will be deformed in expanding matter. It is easy to see
that the radial expansion of the fireball should cause broadening of
the sideward peaks in the $\Delta\phi$--distributions of associated
hadrons. Due to the radial expansion, the peaks will
acquire an additional width of the order of
\mbox{$<\theta_+-\theta_->$}\,. Here $\theta_\pm$ are local values of
the Mach angles in individual events. The angular brackets mean
averaging over the jet trajectory in a given event as well as over all
events with different positions of di--jet axes. Assuming that particle
emission is perpendicular to the surface of Mach cone and taking
$<u>\sim 0.4, c_s\simeq 1/\sqrt{3}$\,, we estimate the angular spread
of emitted hadrons in the range $30^\circ-50^\circ$\,. This is
comparable with the half distance between the away--side peaks of the
$\Delta\phi$ distribution observed by the STAR and PHENIX
Collaborations~\cite{Adl03a,Adl03b,Wan04,Jac05}.  On the basis of this
analysis we conclude that in individual events the sideward maxima
should be asymmetric and more narrow than in an ensemble of
different events. Due to a stronger absorption of particles
emitted from the inner part of the shock (events 1, 3 in 
Fig.~\ref{fig4}), the two peaks may have different amplitudes.   
We think that these effects can be observed by
measuring three--particle correlations (see next section).

There is one more reason for broadening of the 
$\Delta\phi$--distributions which one should keep in mind when comparing
with experimental data: due to the momentum
spread of the initial parton distributions, $\Delta p_*\lesssim 1$\,GeV,
the di--jet system has a nonzero total momentum with respect to the
global c.m. frame.  As a consequence, the angle $\theta_*$
between the trigger-- and the away--side jet is generally--speaking
not equal to $\pi$\,, as was assumed above. Taking typical momenta of
initial partons as $p_0$\,, with $p_0>4-6$\,GeV~\cite{Adl03a,Adl03b,
Wan04,Jac05}, we estimate the angular spread as 
$|\pi-\theta_*|\sim\Delta p_*/p_0\lesssim 0.1$\,. 
Therefore, the considered broadening should be much less than the typical
shift of the Mach angles due to the collective flow.

\section{Influence of longitudinal expansion of QGP}

In the preceding section, only effects of transverse flow have been
considered. On the other hand, experimental data on the jet--induced
pseudorapidity correlations in central Au+Au collisions at RHIC
energies have appeared recently~\cite{Mag04}.  According to these data,
''near--side'' hadrons (emitted in the forward
hemi\-sphere,~$\Delta\phi<\pi/2$\,, with respect to the trigger jet)
exhibit much broader distributions in the relative pseudorapidity,
$\Delta\eta$\,, as compared with  $pp$ and \mbox{d+Au} interactions.
The widths of the $\Delta\eta$ distributions increase with decreasing
$p_T$ of secondary hadrons, reaching values of
\mbox{$\Delta\eta\sim 1$} for $p_T\to 0$\,.

This broadening can be naturally explained as a
consequence of the longitudinal expansion of fireball matter created at
early stages of a heavy--ion collision. Let us consider a trigger jet
emitted at the initial stage of the reaction ($t\simeq 0$) with the
c.m. pseudorapidity $\eta\simeq 0$\,, i.e. in the transverse plane
$z\simeq 0$\,, where $z$ is the beam axis. Below we consider the
Bjorken--type scenario of nuclear collisions~\cite{Bjo83} and
assuming that a cylindrical, longitudinally expanding volume of QGP
with radius $R$ is formed at proper time
$\tau\equiv\sqrt{t^2-z^2}=\tau_0$\,, where
$\tau_0\simeq 0.4-0.6$\,fm/c. In the following discussion we disregard
the radial expansion of the fireball, assuming that the near--side
correlations are formed during a comparatively short
time,~\mbox{$\tau-\tau_0<R/c_s$}\,,which is needed for the trigger jet
to leave the fireball.

Weak, sound--like, perturbations of fireball matter propagate in
$z$--direction with velocities $(u\pm c_s)/(1\pm u\hsp c_s)$\,, where
$u$ and $c_s$ are, respectively, the collective longitudinal velocity
and the sound speed at a given space--time point $(t,z)$\,.
To find the trajectory of a sound wave front
$z=z_+(t)$\,, one should solve the equation for the $C_+$
characteristics~\cite{Lan59}:
\bel{swfe}
\frac{dz_+}{dt}=\left.\frac{u+c_s}{1+uc_s}\right|_{z=z_+(t)}\,.
\ee
Assuming the scaling law, $u=z/t=\tanh{\eta}$\,, for the longitudinal
expansion of the fireball, one can solve~\re{swfe} analytically, even in
the case when $c_s$ is a function of $\tau$\,. Indeed, changing the
variables from $(t,z)$ to $(\tau,\eta)$\,, one gets the following
equation for the pseudorapidity~$\eta_+$ of the sound front
\bel{swf1}
\tau\frac{d\eta_+}{d\tau}=c_s(\tau)\,.
\ee
Assuming that the sound wave is excited at $t=\tau_0\hsp,\,z=0$\,,
we get
\bel{swfs}
\eta_+(\tau)=\int\limits_{\tau_0}^\tau \frac{d\tau}{\tau}\hsp c_s(\tau)\,.
\ee
The expression for the $C_-$ characteristics, $\eta=\eta_-(\tau)$\,, is
given by the r.h.s. of~(\ref{swfs}) with an additional negative sign.
From this derivation we conclude that at fixed proper time
$\tau>\tau_0$ a primary jet, created at $\tau\simeq 0,\,\eta=0$\,,
disturbs the fireball matter in the pseudorapidity
region $\eta_-<\eta<\eta_+$\,. In the case $c_s={\rm const}$ one
obtains a simple expression
\bel{swfs1} \eta_\pm=\pm
c_s\log\left(\frac{\tau}{\tau_0}\right)\,.
\ee

Within this approach one may expect nonzero correlations of secondaries
within the interval of relative pseudorapidities
$\eta_-\leqslant\Delta\eta\leqslant\eta_+$\,.  Substituting
$\tau_0=0.5$\,\,fm/c, \mbox{$\tau=5$\,\,fm/c}, $c_s=1/\sqrt{3}$ we get
an estimate  $|\Delta\eta|_{\rm max}\simeq 1.3$\,. This agrees quite
well with the experimental data of Refs.~\cite{Adl03b,Mag04}
(after subtraction of the ''background'' $pp$ peak from the
experimental $\Delta\phi-\Delta\eta$ distribution).
As one can see from~\re{swfs}, the width of the
$\Delta\eta$ distribution is sensitive to the QGP formation time
$\tau_0$ and, therefore, measurements of these distributions in
different $p_T$ intervals can be used as a clock for the thermalization
process. More detailed information about the $\Delta\eta-\Delta\phi$
asymmetry can be extracted from 3--particle correlations. In
particular, one can use such correlations to select events where the
trigger and two associated hadrons have momenta in the axial plane
(containing the beam axis). If the pseudorapidity of the trigger jet is
small, both associated hadrons will probe the longitudinal flow of the
QGP.

In the case of a strong shock we expect that the matter in the
inner part of the Mach region will be swept to its
surface\hsp\footnote{
Strong shocks produced by a localized, high energy initial
perturbation in the QGP have been studied
by using the 3D fluid--dynamical model in Ref.~\cite{Dum05}.
Clear deviations from Mach cone shape have been
predicted.
}.
This will result in a local depletion of parton
density inside the Mach cone. Such a behavior should be observed as a
dip in pseudorapidity density of the associated hadrons around
$\eta=0$\,. It seems that the experimental data on near--side
production~\cite{Adl03b} indeed show such a dip for soft secondaries.

\section{Conclusion}

In this paper we have investigated properties of Mach
shock waves induced by high--energy partons
propagating through dense quark--gluon matter
created in heavy--ion collisions.
By using simple kinematic and hydrodynamic relations
we have analyzed deformations of these shocks
due to the radial and longitudinal expansion of QGP.
Taking typical flow parameters expected in central collisions
of nuclei at RHIC and LHC energies we show
that the shape and orientation of Mach regions
are strongly modified as compared to the case of static
(nonexpanding) medium.
This may obscure observable signatures of Mach collective waves.

In the future we are going to take into account several additional
effects, in particular, bending of the leading parton's trajectory due
to the friction force from moving medium. In principle, this bending
may also produce isolated peaks in the away--side $\Delta\phi$
distributions. On the contrary, the Mach waves should generate
ring--like maxima around the jet axis. We believe
that measurements of three--particle correlations with a broad coverage
of angular space can be used to differentiate between these two
possibilities. We are also planning to investigate properties of strong Mach
shocks, in particular, their sensitivity to the equation of state of
QGP and to effects of collective expansion.

\begin{acknowledgments}
The authors thank  A. Dumitru, M.I. Gorenstein,
D.H. Rischke and E.V. Shuryak for
useful discussions. This work has been supported by the GSI,
the DFG grant 436 RUS 113/711/0-2 (Germany),
the RFBR Grant 03--02--04007, and the MIS Grant
NSH--1885.2003.2 (Russia).
\end{acknowledgments}


\end{document}